# Seven properties of self-organization in the human brain

**Birgitta Dresp-Langley** [1]*

[1]  Centre National de la Recherche Scientifique (CNRS) France, ICube UMR 7357 CNRS-University of Strasbourg; birgitta.dresp@unistra.fr

*  Correspondence: birgitta.dresp@unistra.fr;

**Abstract:** The *principle of self-organization* has acquired a fundamental significance in the newly emerging field of computational philosophy. Self-organizing systems have been described in various domains in science and philosophy including physics, neuroscience, biology and medicine, ecology, and sociology. While system architecture and their general purpose may depend on domain specific concepts and definitions, there are (at least) seven key properties of *self-organization* clearly identified in brain systems: 1) modular connectivity, 2) unsupervised learning, 3) adaptive ability, 4) functional resiliency, 5) functional plasticity, 6) from-local-to-global functional organization and 7) dynamic system growth. These are defined here in the light of insight from neurobiology, cognitive neuroscience and Adaptive Resonance Theory (ART), and physics to show that self-organization achieves stability and functional plasticity while minimizing structural system complexity. A specific example informed by empirical research is discussed to illustrate how modularity, adaptive learning, and dynamic network growth enable stable yet plastic somatosensory representation for human grip force control. Implications for the design of "strong" artificial intelligence in robotics are brought forward.

## 1. Introduction

The principle of *self-organization* [1] governs both structure and function, which co-evolve in self-organizing systems. Self-organizing systems [2] differ from any computational system where the architecture and all its functional aspects are created and controlled by their designer. In line with previous attempts at a comprehensive definition of the concept [1], the author proposes that *self-organization* may be defined in terms of a general principle of functional organization that ensures a system's auto-regulation, stability, adaptation to new constraints, and functional autonomy. A self-organizing system [2], beyond the fact that it regulates and adapts its own behavior [2,3], is capable of creating its own functional organization [2,3,4]. The concept of *self-organization* acquires a critically important place in the newly emerging field of computational philosophy, where it inspires and lends conceptual support to new approaches to complex problems, in particular in the field of Artificial Intelligence (AI) [5].

Mathematical developments evoking *self-organization* as a general functional principle of learning and adaptation in biological systems hark back to Hebb's work on synaptic plasticity [6], and to work by Minsky and colleagues [7] at the dawn of research on artificial intelligence in the context of Rosenblatt's PERCEPTRON model [8]. *Self-organization* is the foundation of what is sometimes referred to as "strong AI" [5]. Systems with self-organizing properties [2] have been developed in physics [9], ecology and sociology [10,11], biology and medicine [12], and in neuroscience [3,13] and perceptual neuroscience [3,14] in continuity with the earlier PERCEPTRON approaches. Structure and functional organization of self-organizing systems vary depending on the field. Their properties relate to function more than to components. The fields of neuroscience and artificial intelligence in particular share a history of interaction in the theoretical development of both the concept of *self-organization* and self-organizing systems, and many of the current advances in AI were inspired by the study of neural processes in humans and other living species [3,5,13].

Neuroscience provides a source of inspiration for new algorithms and architectures, independent of and complementary to mathematical methods. Such inspiration is well-reflected by many of the concepts and ideas that have largely dominated traditional approaches to AI. Neuroscience may also convey external validity to AI. If an algorithm turns out to prove a good model for a functionally identified process or mechanism in the brain, then such biological plausibility lends strong support to the fitness of the algorithm for the design of an intelligent system. Neuroscience may thus help conceive new algorithms, architectures, functions, and codes of representation for the design of biologically plausible AI by using a way of thinking about similarities and analogies between natural and artificial intelligence [15]. Such two-way conceptual processes acquires a particular importance in the newly emerging field of computational philosophy, which regroups a wide range of approaches relating to all fields of science.

Computational philosophy [16] is aimed at applying computational techniques, models, and concepts to advance philosophical and scientific discovery, exploration, and argument. Computational philosophy is neither the philosophy *of* computation, an area that asks about the nature of computation itself, nor the philosophy *of* artificial intelligence. Computational philosophy represents a self-sufficient area with a widespread application across the full range of scientific and philosophical domains. Topics explored in computational philosophy may draw from standard computer programming, software engineering, artificial intelligence, neural networks, systems science, complex adaptive systems, and computer modeling. As a relatively young and still growing domain, its field of application is broad and unrestricted within the traditional discipline of general philosophy. In the times of Newton, there was no epistemological boundary between philosophy and science. Across the history of science, there has never been a clear division between either computational and non-computational philosophy, or computational philosophy and other computational disciplines [16].

The place of computational philosophy in science entirely depends on the viewpoint adopted, and goal pursued by the investigator [17,18]. If the goal pursued is to enrich computational philosophy based on an understanding of brain processes, then the functional characteristics of brain mechanisms may fuel the development of computational philosophy. If the goal is to enrich brain science based on computational philosophy, then empirical brain research will be fueled by computational philosophy for building brain models reflective of mechanisms identified in the human brain [17,18]. This article is a conceptual essay written from the viewpoint of computational philosophy. It highlights seven general functional key properties related to the principle of *self-organization*, which are then discussed under the light of a specific example from sensory neuroscience, backed by empirical data. How modularity, adaptive learning, and dynamic network growth enable stable somato-sensory representation for human grip force control in a biological neural network (*from hand to brain and back*) with previously identified functional plasticity is illustrated.

Since structure and functional organization co-evolve in self-organizing systems [1,2], one cannot account for such systems without providing an account for the functional properties most closely linked to its self-organizing capacity. The latter is generally described in terms of spatiotemporal synergies [1,9]. In the brain, neurons respond at time scales of milliseconds, while perception, which is experience and memory dependent, takes longer to form. Such timescale separation between long-time scale parameters and short-time scale functioning [19,20] is akin to that described for physical synergetic systems [9], and reflects the *circular causality* that is characteristic of *self-organization* in general [1]. The following sections start with an overview of seven key properties of systems that "self-organize". This is followed by a discussion of examples of such properties in the human somato-sensory system [21,22] involved in the control of prehensile synergies for grip-force adaptation. The example provides a biologically plausible conceptual support for the design of autonomous *self-organization* (AI) in soft robotics, and illustrates why the seven key properties brought forward here in this concept paper are conducive to advancing the development of "strong AI" [5], as pin-pointed in the conclusions.

## 2. Seven key properties of *self-organization*

Seven properties linked to the principle of *self-organization* have been described on the basis of functional investigation of the human brain: 1) modular connectivity, 2) unsupervised learning, 3) adaptive ability, 4) functional resiliency, 5) functional plasticity 6) from-local-to-global functional organization and 7) dynamic system growth.

### 2.1. Modular functional architecture and connectivity

Modularity refers to a computational and/or structural design principle for systems that can be decomposed into interacting subsystems (nodes, modules) that can be understood independently. Modular systems design is aimed at reducing complexity [23] by a fundamental design principle identified in biological neuronal systems at the scale of cells (units, neurons), local circuits (nodes), and interconnected brain areas (subsystems) [24]. The human brain's neuronal network architecture is not based on a genetically preformatted design, although some of it may be prewired, but is progressively shaped during ontogenetic development by physiological and chemical changes that obey computational rules of activity-dependent *self-organization* [25]. At the medium level of local circuits, the brain (cortex) is organized in local clusters of tightly interconnected neurons that share common input. The neuronal targets that constitute a basic computational module share similar functional properties. Activity-dependent *self-organization* influences the system's modularity on the one hand, and modular connectivity promotes spontaneous firing activity on the other [25]. Thus, the modular connectivity of a self-organizing system and its capacity of *self-organization* are interdependent, and they co-evolve in a mutually reinforcing process to ensure the simultaneous development of both structural and functional capacity. This entails that the more such a system learns, the more active connections it will develop on the one hand, and the more it will be able to learn, on the other. For its structural and functional development, a self-organizing system exclusively uses unsupervised learning.

### 2.2. Unsupervised Learning

Unsupervised learning [6,26,27,28], one may also call it self-reinforced learning, is essential to the principle of self-organization, as illustrated by functional dynamics of the neural network systems described in Adaptive Resonance Theory (ART) and Self-Organizing Maps (SOM). Unsupervised learning is essential to overcome the stability–plasticity dilemma in neural networks, and both ART [27] and SOM [28,29] are based on unsupervised approaches that are fundamental in machine learning in general, and in Artificial Intelligence (AI) in particular. The *Hebbian* synapse and synaptic learning rules [6] are the fundamental conceptual basis of unsupervised learning in biological [30] and artificial neural networks [31]. A synapse refers to connection between two neurons in a biological or artificial neural network, where the neuron transmitting information via a synapse or synaptic connection is referred as the *pre-synaptic neuron*, and the neuron receiving the information at the other end of a synaptic connection as the *post-synaptic* neuron (Figure 1). The information propagation efficiency of biological and artificial synapses is strictly self-reinforcing as the more a synapse is stimulated, the more effectively information flows through the connection, which ultimately results in what Hebb [6], and subsequently others [3,31] have called Long-Term Potentiation (LTP). Synaptic connections that are not repeatedly stimulated and as a consequence not self-reinforced will lose their information propagation efficiency, which ultimately results in Long-Term Depression (LTD).

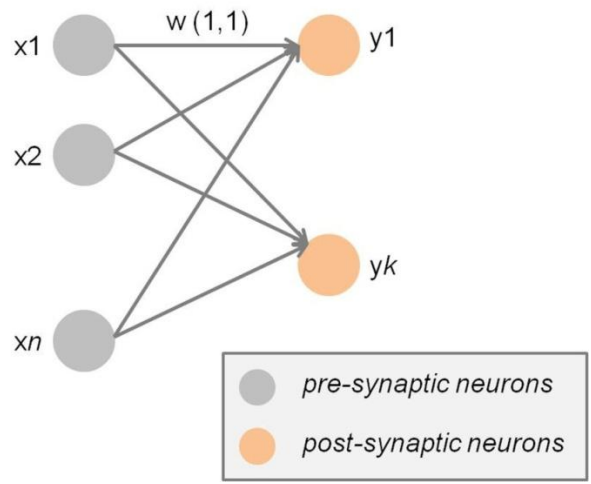

**Figure 1:** Schematic illustration of Hebbian synapses within a small neural network. Self-reinforcing synaptic learning is by definition unsupervised and involves the progressive increment of the synaptic weights (w) of efficiently stimulated connections, which are thereby long-term potentiated, while non-reinforced synapses will lose their efficiency and ultimately become long-term depressed.

The information propagation in unsupervised synaptic learning in neural networks may be event-driven [32], clock-driven [33], or a combination of both [31]. The Long-Term Potentiation (LTP) of efficient synaptic connections on the one hand, and the Long-Term Depression (LTD) of inefficient connections on the other promote the emergence of an increasingly effective functional organization in the neural network akin to that found in biological organisms, chemical structures, and ecosystems. In computational thinking and philosophy, self-reinforcing Hebbian learning may, indeed, be seen as a ground condition for adaptive system function, a highly dynamic, unsupervised learning process where local connections change state towards potentiation or depression, depending on how efficiently they propagate information across the system. Synaptic long-term potentiation and long-term depression are not definitive. A depressed state may reverse to a potentiated one, and *vice versa*, as a function of a persistent change in the system's external environment (stimuli), or of a specific chemical treatment or drug in the case of biological neural networks. LTP and LTD have acquired a potentially important role in contemporary neuroscience [34], and may be exploited to treat disorder and disease in the human brain, knowing that a variety of neurological conditions arise from either lost, or excessive synaptic drive due to sensory deprivation during childhood, brain lesions, or disease [35]. Manipulation of relative synaptic efficiency using various technologies may provide a means of normalizing synaptic strength and thereby ameliorating plasticity-related disorders of the brain [34,35]. Thus it may, indeed, be argued that the reversibility of synaptic efficiency as a function of changes during self-reinforcing learning drives the self-organizing system's adaptive ability.

### 2.3. *Adaptive ability*

A self-organizing system will adapt its functional organization to significant changes in the environment by the ability to learn new "tricks" to cope with new problems of increasing complexity. Biological neural networks in the human central nervous system have the ability to adapt their functional fine-tuning to sudden changes in the environment [36], making the brain remarkably efficient at coping with an often unpredictable, ever-changing external world. In any self-organizing neural network, such adaptation relies heavily on their modular connectivity on the one hand, and on synaptic (*Hebbian*) plasticity, on the other [37,38,39]. Research on so-called neuromodulation [40,41,42], a biological mechanism that dynamically controls intrinsic properties of neurons and their response to external stimuli in a context-dependent manner, has produced simulations of adaptive system behavior. Adaptive Resonance Theory (ART) uses *self-organization* to explain how normal and abnormal brains learn to categorize and recognize objects and events in a changing world, and how normal as well as abnormal learned categories may be consolidated in memory for a long time [42]. Thus, brain pathology may be conceived in terms of a self-organizing system's adaptive ability gone wrong. This possibility was already to some extent taken into consideration by Darwin [43], who noted that natural selection can act together with other processes, including random changes in the frequencies of phenotypic differences that are not under strong selection, and changes in the environment, which may reflect evolutionary changes in the organisms themselves. At the cellular level, adaptation refers to changes in a cell's or neuron's behavior in response to adverse or varying environmental changes. Adaptation may be normal, or pathological (abnormal), depending on extent and type of environmental pressure on the cell, a system of cells, or a whole brain network [44,45,46,47]. Under extreme external conditions, some forms of pathological adaptation may be an effective means towards the general goal of survival. In human behavior, the Stockholm Syndrome is one such example, where hostages start taking sides with their aggressors to functionally adapt to the terrible fact that they are at their complete mercy. Adaptive system ability as a concept, in the brain sciences and in computational philosophy, helps conceive intelligent systems with a capacity to generate order from, or preserve order within, external chaos. Seemingly random perturbations will help a self-organizing system develop and perform even better, rather than prevent its evolution. Perturbations will promote the emergence of an increasingly effective functional organization, as found in biological organisms, chemical structures, or ecosystems. *Self-organizing* adaptation is to be seen as a highly dynamic process [20,46,47], where components are constantly changing state as a function of state changes in other components.

Such complex mutual dependency in self-organization was not known in the times of Darwin, and is therefore not included in traditional definitions of the concept "adaptation" [43,44,45]. The adaptive systemic changes we are talking about here in this concept paper are determined by self-reinforcement of connections that profit the system's functioning and ensure its dynamic functional growth on the one hand, and by local suppression of connections that are either redundant, or disserve the system, on the other. Both processes, reinforcement and inhibition, are critical to sustain a self-organizing system's ability to cope with unexpected external changes or pressure, and thereby also ensure its functional resiliency.

### 2.4. Functional resiliency

A self-organizing system's adaptive ability implies functional resiliency. After lesion or damage, a human brain will continue to function, often astonishingly well and without any detectable change in efficiency. The human brain can endure numerous micro-strokes with seemingly no detrimental impact, and is resilient against both targeted and random damage or lesions [48,49]. Self-organizing systems, like the human brain, are intrinsically robust and can withstand a variety of perturbations, even partial destruction. The strength of functional interaction between any two system nodes is not solely determined by the presence or absence of a direct connection, but mostly by the number of indirect (long-range) connections [50]. These long-range connections, which will be discussed in greater detail in 5) here below, are indispensable to ensure self-repair or self-correction of partial systemic damage, and make the system capable of returning to its initial functional state after local damage. By virtue of their modular connectivity discussed here above in 1), and self-reinforcing learning capacity discussed here above in 2), system components or subsystems (synapses or networks) that have initially learnt to fulfill a specific function can spontaneously adapt to perform a different, new function that was previously ensured by the damaged component(s). This self-organizing ability is referred to as functional plasticity.

### 2.5 Functional plasticity

The functional resiliency of a self-organizing system implies functional plasticity [21,22,51], which is a necessary ground condition for system resiliency, but also achieves a purpose well beyond. Functional plasticity ensures system functioning under adverse conditions and/or after partial system damage. Posttraumatic stress disorder (PTSD), for example, is associated with plastic functional changes in the human medial prefrontal cortex, hippocampus, and amygdala that correlate with a smaller hippocampal volume, and both reversed to normal after treatment [52]. Like in the human brain, where a functional subsystem may take over the functions of another after brain damage [21,22,51], a functional subsystem may appear spontaneously and maintain its function autonomously by *self-organization* in a computer generated system. The control needed to achieve this has to be distributed across system levels, components or cells, and/or sub-systems. If system control were centralized in a subsystem or module, then the system as whole would lose its organization whenever the sub-system is damaged or destroyed. Use-dependent long-term changes of neuronal response properties must be gated to prevent irrelevant activity from inducing inappropriate modifications. Local network dynamics contribute to such gating, as synaptic modifications [53] depend on temporal contiguity between pre-synaptic and post-synaptic activity, there are observable stimulation-dependent modifications, as shown on the example of orientation selectivity in adult cat visual cortex [35]. The stability-plasticity dilemma, a constraint for intelligent systems, is potentially resolved in self-organizing systems, such as those in ART [3,14,47]. Plasticity is necessary for the integration of new knowledge by self-reinforced learning, but too much of it compromises systemic stability and may cause catastrophic forgetting of previous knowledge [54]. It is assumed that too much plasticity will result in previously learnt data being constantly forgotten, whereas too much stability will hinder self-reinforce learning at the synaptic level, yet, the exact functional relationship between changes in synaptic efficacy and structural plasticity is not entirely understood. It has been proposed that a continuum exists between the two, such that changes in synaptic efficacy precede and instruct structural changes, however, in other cases, structural changes may occur without any stimulation producing an initial change in local synaptic efficiency [55], which points towards the critical functional role of long-range connections [56] within the from-local-to-global functional organization of self-organizing systems.

### 2.6.*From-local-to-global functional organization*

In a self-organizing neural network, changes in the system during self-reinforced synaptic learning are initially local, as components or neurons initially only interact with their nearest "neighbors". Though local connections are initially independent of connections farther away, *self-organization* generates "global order" on the basis of many spontaneous, initially local, interactions [56], where the most efficient synaptic connections self-reinforce, are long-term potentiated as described here above in 2) and, ultimately, acquire propagation capacity beyond local connections. This leads to the formation of functionally specified long-range connections, or circuits which, by virtue of *self-organization*, self-reinforce on the basis of the same *Hebbian* principles that apply to single synapses, however, the rules by which long-range circuits of connections learn can no longer be accounted for in terms of a *Hebbian* linear model. The human brain is, again, the choice example of a complex biological structure where local, modular processing potentiates global integrative processing. Current functional brain anatomy suggests areas that form domain-specific hierarchical connections [57,58] on the one hand, and multimodal association areas receiving projections from more widely distributed functional subsystems [59]. Dominance of one connectivity profile over the other can be identified for many areas [56], revealing the self-organizing principles of long-range cortical-cortical functional connectivity. Early visual cortical areas such as V1 and V2 already show a functional organization beyond strictly local hierarchical connections [58,60]. The prefrontal, temporal, and limbic areas display "functional hubs" [56], projecting long-range connections across larger distances to form the "neural epicenters" [56] of scale-free, distributed brain networks [61,62]. The "beyond the classic receptive field" functional organization of the visual brain was progressively unraveled in behavioral and functional neuroscience over the last 30 years [60]. The discovery of input effects from beyond the "classic receptive field", as previously identified and functionally defined in much earlier, Nobelprize awarded work [63-66], has shown that neuronal activity recorded from cortical areas V1 and V2 in response to visual stimuli is modulated by stimuli presented outside the corresponding receptive fields on the retina [67-69]. This is direct evidence for contextual modulation of neural activity, and indirectly reflects functional properties of long-range neural connections at early processing levels in the visual brain [68,69]. In higher visual areas such as the temporal lobe, visual receptive fields increase in size and lose retinotopic organization, encoding increasingly complex features [60]. This from-simple-to-complex, self-organizing functional hierarchy forms the core of the LAMINART model family [70-71] of Adaptive Resonance Theory [3,14,47]. In the LAMINART neural networks, self-organizing long-range cooperation and short-range competition, whereby locally stimulated bipolar neurons complete boundaries across gaps in oriented line or edge contrast stimuli by receiving strong excitatory inputs from both sides, or just one side of their receptive fields. The more strongly activated bipolar cells inhibit surrounding bipolar cells within a spatially short-range competitive network. The short-range network communicates with long-range resonant feedback networks connecting the interblob and blob cortical streams within V1, V2, and V4 of the visual cortex. The resonant feedback networks enable boundaries and surfaces in images to emerge as consistent representations on the basis of computationally complementary rules. This self-organizing property of resonant feedback networks in ART is called complementary consistency [72]; the computational mechanisms that ensure complementary consistency contribute to three-dimensional perceptual organization [72,73,74,75]. The long-range resonant properties of the neural network architectures exploited by ART enable the self-organizing system to grow dynamically.

*2.7. Dynamic functional growth*

A self-organizing system is dynamic and its components (cells, neurons, circuits) are constantly changing states relative to each other. As explained here above under 1), structure and function of the system are mutually dependent, which entails that the changes that occur while such a system is developing further, i.e. growing, are not arbitrary but activity-dependent [76,77,78,99]. While the system grows by changing states, there will be relative states that will be particularly beneficial to the system's effectiveness and, as a consequence, these states self-reinforce along similar principles as those described here above in 2). When consistently reinforced, newly emerging system states will, ultimately, become stable states, but with functional plasticity as explained here above in 4) to resolve the stability-plasticity dilemma [3,54]. Less beneficial or useless new relative states will not self-reinforce and, as a consequence, be inhibited and ultimately functionally depressed. Each connection within a self-organizing system has its own, individual characteristics, like a species within an ecosystem. A particular example of non-linear dynamic functional growth in fixed-size neural networks (Figure 2) would be activity-dependent formation of dedicated resonant circuitry [3,79].

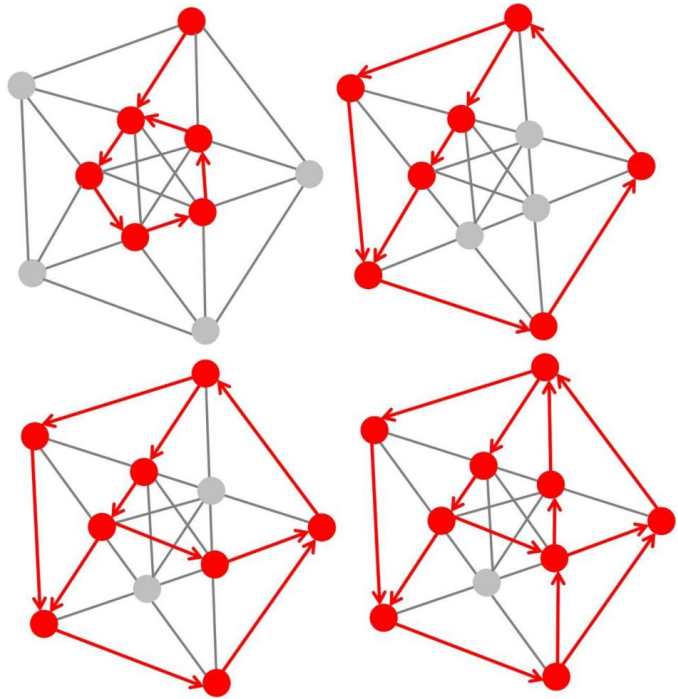

**Figure 2:** Schematic illustration of self-organizing formation of dedicated resonant circuitry in a fixed-size neural network of ten neurons only. An exponential number of functional states are possible therein, allowing even a small-size network to develop new, dynamic functionalities without the need for adding further structural complexity. Resonant neurons (highlighted in red colour) are primed throughout their functional development to preferentially process input which carries statistically "strong" signals, as explained previously in [79]. When activated, resonant neurons send signals along all delay paths originating from them, and all those receiving a signal coinciding with the next input signal remain activated. The formation of resonant circuitry is activity-dependent; long-term potentiated connections between resonant neurons may become progressively depressed, as in the model of self-reinforcing (*Hebbian*) synaptic learning, when their function is no longer activated.

The amount of different individual characteristics within the functional structure directly determines the system's functional complexity. Individual components, or cells, fit in a functionally specified "niche" within the system. The propagation of fits is self-reinforcing by nature, and the larger the niche, the quicker the propagation of additional functions in increasingly larger sub-circuits, exerting increasingly stronger attraction on functionally still independent cells. Such propagation of fits drives a positive feedback process enabling explosive system growth, and the production of new functional circuitry in existing network structures. A system's growth generally stops when the system resources are exhausted. A self-organizing system, however, never stops growing dynamically. As illustrated here above on the example of resonant sub-circuitry formation in structurally fixed/limited neural networks (Figure 2), self-organized functional growth does not require adding structural component (cells, neurons, subsystems). The system can grow and develop new, dynamic functionalities without the need for adding further structural complexity. As the external environment of the system changes, functional components or cells directly interacting with the environment will adapt their state(s) in a self-reinforcing process to maintain their fitness within the system. This adaptive fit will propagate further inwards, until the whole functional structure is fully adapted to the new situation. Thus, a dynamically growing self-organizing system constantly re-organizes by mutually balancing internal and external pressures for change while trying to maintain its general functional organization, and to counteract any loss thereof. Functional self-preservation is, indeed, a self-organizing system's main purpose, and each component or cell is adaptively tuned to perform towards this goal. A self-organized system is stable, largely scale-invariant, and robust against adverse conditions. At the same time, it is highly dynamic. In physics, systems achieve so-called "criticality" by the fine-tuning of control parameters to a specific value for which system variations become scale-invariant. In biological systems, criticality occurs without the need for such fine-tuning. The human brain is an example of such a system. This self-tuning to criticality, accounted for in physics by graph theory, is called *self-organized criticality* [80].

## 3. Seven properties *of self-organization* in a somatosensory neural network

The brain structures which subserve cognitive functions require sensory experience for the formation of neuronal connections by self-organization during learning [81]. Neuronal activity and the development of functionally specific neural networks in the continuously learning brain are thus modulated by sensory signals. The somatosensory cortical network [82], or S1 map, in the primate brain is an example of such self-organization. S1 refers to a neocortical area that responds primarily to tactile stimulations on the skin or hair. Somatosensory neurons have the smallest receptive fields and receive the shortest-latency input from the receptor periphery. The S1 cortical area is conceptualized in current state of the art [82,83] as containing a single map of the receptor periphery. The somatosensory cortical network has a *modular functional architecture and connectivity (property 1)*, with highly specific connectivity patterns [82,83,84,85], binding functionally distinct neuronal subpopulations from other cortical areas into motor circuit modules at several hierarchical levels [84]. The functional modules display a hierarchy of interleaved circuits connecting via inter-neurons in the spinal cord, in visual sensory areas, and in motor cortex with feed-back loops, and bilateral communication with supraspinal centers [84,85]. The *from-local-to-global functional organization (property 6)* of motor circuits relates to precise connectivity patterns, and these patterns frequently correlate with specific behavioral functions of motor output. Current state of the art suggest that developmental specification, where neuronal subpopulations are specified in a process of precisely timed neurogenesis [85], determines the *self-organizing* nature of this connectivity for motor control, in particular limb movement control [84,85]. The *functional plasticity (property 5)* of the somatosensory cortical network is revealed by neuroscience research investigating the somatosensory cortical map has shown that brain representations change adaptively following digit amputation in adult monkeys [21]. In the human primate [86], somatosensory representations of the fingers left intact after amputation of others on the same hand become expanded in less than ten days after amputation, when compared with representations in the intact hand of the same patient, or to representations in either hand of controls. Such network expansion reflects the *functional resiliency (property 4)* of the self-organized somatosensory system.

The human hand has evolved [87] as a function of active constraints [88-100], and in harmony with other sensory systems such as the visual and auditory brain [97,99,101]. Grip force profiles are a direct reflection of complex low-level, cognitive, and behavioral synergies this evolution has produced [87-101]. The state of the art in experimental studies on grip force control for lifting and manipulating objects [88,89,91,93,94] provides insight into the contributions of each finger to overall grip strength and fine grip force control. The middle finger, for example, has evolved to become the most important contributor to gross total grip force and, therefore, is most important for getting a good grip of heavy objects to lift or carry, while the ring finger and the small (pinky) finger have evolved for the fine control of subtle grip force modulations, which is important in precision tasks [102-106]. Human grip force is governed by self-organizing prehensile synergies [91,92] that involve *from-local-to-global functional interactions (property 6)* between sensory (low-level) and central (high-level) representations in the somatosensory brain. Grip force can be stronger in the dominant hand compared with the non-dominant hand, and may reverse spontaneously depending on the necessity for *adaptive ability (property 3)* as a function of specific environmental constraints [95,96,100]. In recent studies, the grip force profiles from thousands of force sensor measurements collected from specific locations on anatomically relevant finger parts on the dominant and non-dominant hands revealed spontaneous adaptive grip force changes in response to sensory stimuli [105], and long-term *functional plasticity (property 5)* as a function of task expertise [103,104].

Somatosensory cortical neural networks of the S1 map [81] develop their functional connectivity [83-86] through a self-organizing process of activity-dependent, *dynamic functional growth (property 7)*. This process is fueled by *unsupervised learning (property 2)* and, more specifically, synaptic (cf. *Hebbian*) learning, which drives spontaneous functional adaptation as well as long-term functional re-organization and plasticity [22,23,54,56,83]. An example of this process, fueled by recent empirical data from thousands of sensor data collected from anatomically relevant locations in the dominant and non-dominant hands of human adults, is illustrated here in Figures 3, 4 and 5.

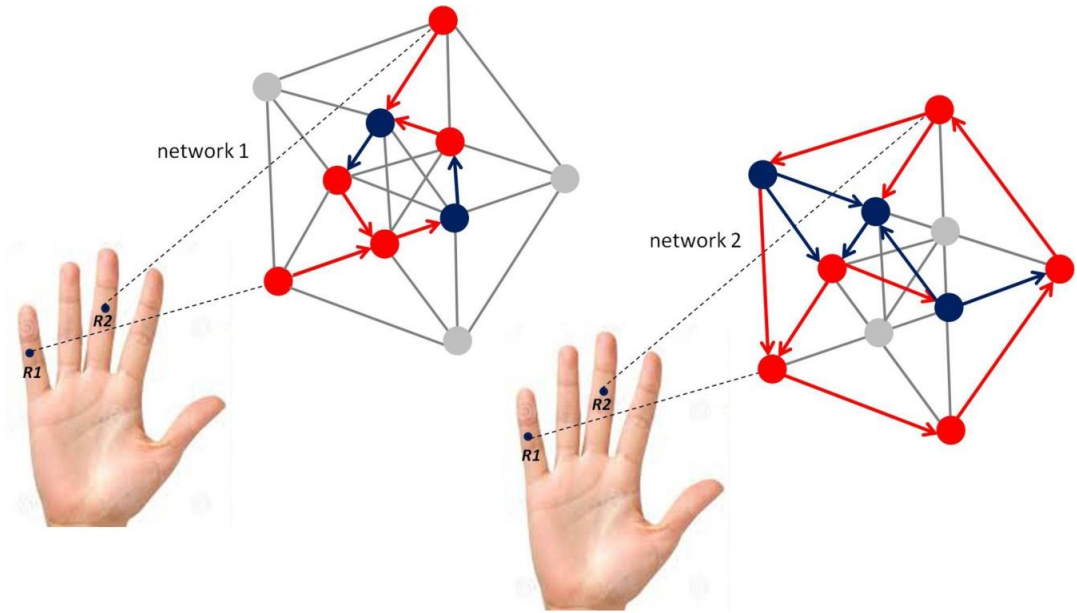

**Figure 3:** Schematic illustration of self-organizing functional reorganization in a fixed-size neural network in the somatosensory brain before (left) and after (right) unsupervised learning of a visually guided manual robotic precision task [103,104]. Mechanoreceptors on the middle phalanxes of the small (R1) and the middle fingers (R2) are indicated. Two different network representations correspond to brain-behaviour states before (left) and after acquisition of grip force expertise (right) for performing the robotic precision task. The different levels of connectivity used here are arbitrarily chosen, and for illustration only; single nodes in the networks displayed graphically here may correspond to single neurons, or to a subpopulation of neurons with the same functional role. Red nodes may represent motor cortex (M) neurons, blue nodes may represent connecting visual neurons (V). Only one-way propagation is shown here to keep the graphics simple, knowing that the somatosensory brain has multiple two-way propagation pathways with functional feed-back loops [82-85].

The grip force profiles of a novice (beginner) and a skilled expert in the manipulation of the robotic device directly reflect such differences in somatosensory cortical representation before and after learning. Individual grip force profiles, corresponding to thousands of individual force sensor data, were recorded in real time from different sensor locations, including R1 and R2 (Figure 3) on the middle phalanxes of the small (R1) and the middle fingers (R2) in the hands of a total beginner and the expert across several robotic task sessions. The grip force profiles are shown in Figure 4 here below. They display two radically different functional states with respect to gross (middle finger) and fine (small finger) grip force control, translating motor expertise and functionally reorganized somatosensory network states driven by self-organization.

The self-organized functional reorganization due to plasticity shown here is stable, as reflected by stable grip force profiles in the expert across task sessions [103]. By comparison, the grip force profiles of the beginner do not display the same stability, as shown by the statistical analyses reported elsewhere [103,104]. It may be assumed that stable dynamic functional growth in the neural network system generating the somatosensory representations is driven by such long-term plasticity, which reflects a process of long-term adaptation to specific task constraints.

This long-term adaptation ensures system stability, but does not reflect a permanent system state. However, the somatosensory system also displays spontaneous functional plasticity and reorganization, and rapid adaptation to new constraints [21,22,81-85].

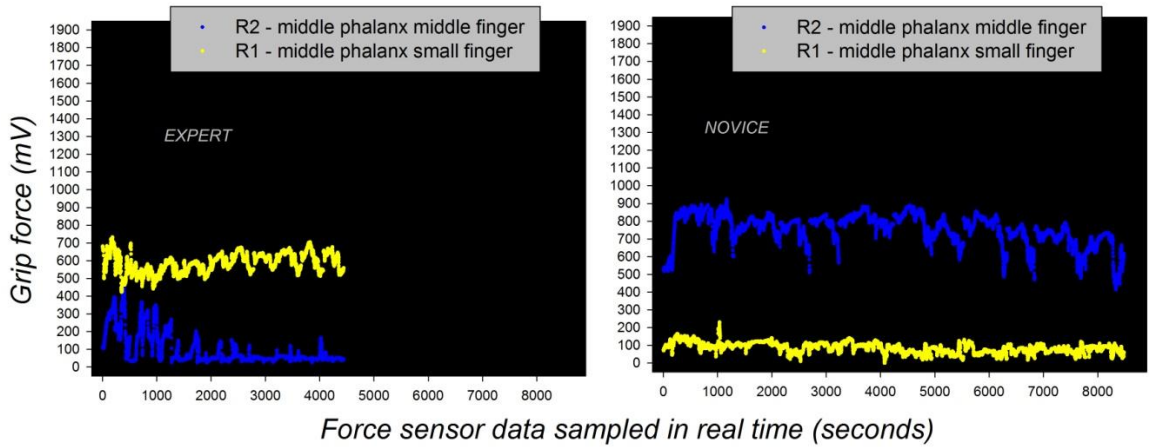

**Figure 4:** Individual grip force profiles of an expert (left) and a beginner (right) in a robotic task [103,104] reflecting two radically different functional states with respect to gross (middle finger) and fine (small finger) grip force control, translating motor expertise and functionally reorganized somatosensory network states (cf. Figure 3), governed by self-organization.

The grip force profiles of one and the same individual adapt spontaneously to new sensory input from other modalities, as predicted by the general functional organization of the somatosensory brain networks [82-86]. Such spontaneous adaptive ability is reflected by dynamic changes that are short-term potentiated, rather than reflective of long-term plasticity. An example of spontaneous grip force adaptation to new visual input in one and the same individual is shown here below in Figure 5. The subject was blindfolded first, then made to see again, during a bimanual grip task where young male adults had to move two weighted handles up and down [105]. The individual grip force profiles corresponding to force sensor recordings from the middle phalanxes of the forefinger and the middle finger in the dominant hand are shown for comparison (Figure 5).

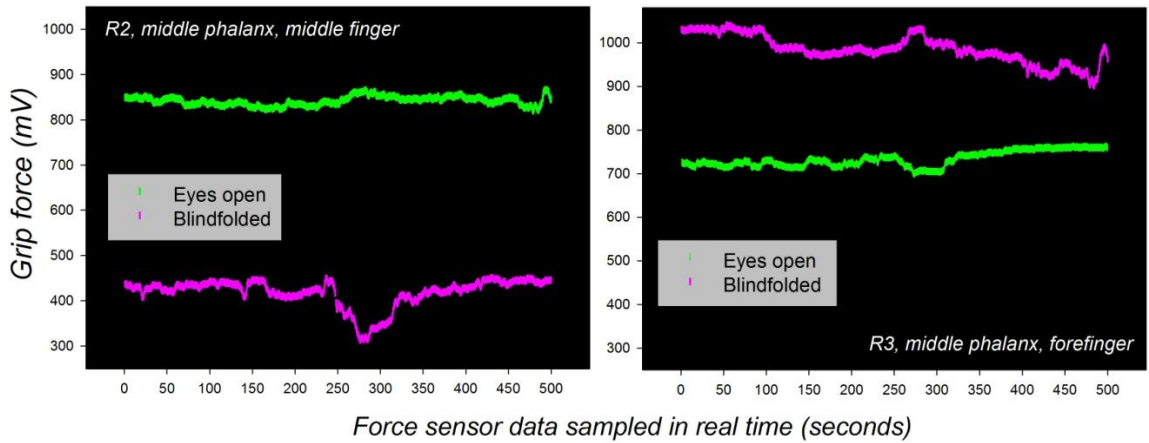

**Figure 5:** Spontaneous adaptive reorganization of an individual's grip force profile (dominant hand) during a bimanual grip task executed first blindfolded, and then with both eyes open. Unpublished data from a study described in [106] are shown. Gross grip force in the middle finger spontaneously increases with sudden visual input (left), while simultaneous grip force in the forefinger decreases (right).

The example here above illustrates seven key properties of self-organization in the somatosensory brain, and points towards the implications of self-organized brain learning for the design of robust control schemes in large complex systems with unknown dynamics, which are difficult to model [108]. Beyond the functional stability and resilience of self-organizing systems, self-reinforced unsupervised learning based on differential plasticity, with feedback control through internal system dynamics, may enable robots [107] to learn to relate objects on the basis of specific sensorimotor representations, for example.

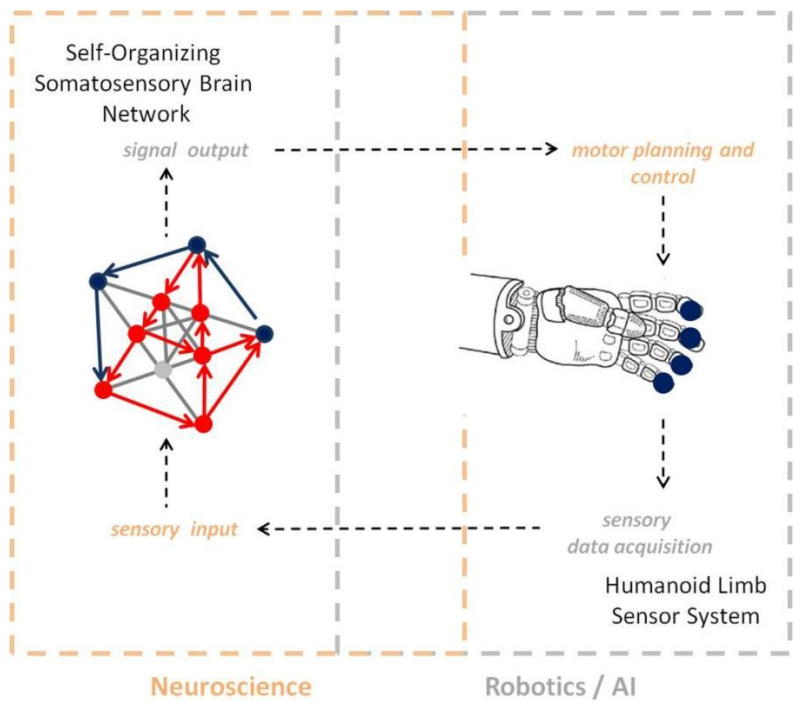

**Figure 6:** Conceptual diagram illustrating a "closed-loop" link between systems neuroscience and robotics / AI.

**Conclusions**

*Self-organization* is a major functional principle that allows us to understand how the neural networks of the human brain continuously generate new knowledge at all hierarchical levels, from sensory to cognitive representation. From the philosophical standpoint, the principle of *self-organization* establishes a clear functional link between the "mental" and the "physical" [109,110]. As a fundamental conceptual support for the design of intelligent systems, it provides conceptual as well as mathematical [1,2,3] tools to help achieve system stability and reliability, while minimizing system complexity. It enables specific fields such as robotics to conceive new, adaptive solutions based on self-reinforced systemic learning, where the activity of connections directly determines their performance and, beyond robotics, allows for the conceptual design of a whole variety of adaptive systems that are able to learn and grow independently without the need for adding non-necessary structural complexity. There is a right balance between structural and functional complexity, and this balance conveys functional system plasticity; adaptive learning allows the system to stabilize but, at the same time, remain functionally dynamic and able to learn new data. Activity-dependent functional systemic growth in minimalistic sized network structures is probably the strongest advantage of *self-organization*; it reduces structural complexity to a minimum, and promotes dimensionality reduction [111,112], which is a fundamental quality in the design of "strong" Artificial Intelligence [5]. Bigger neural networks akin to those currently used for deep learning [113] do not necessarily learn better or perform better [114]. *Self-organization* is the key to designing networks that will learn increasingly larger amounts of data increasingly faster as they learn, consolidate what has been learnt, and generate output that is predictive [47,68] instead of being just accurate. In this respect, the principle of *self-organization* will help design Artificial Intelligence that is not only reliable, but also meets the principle of scientific parsimony, where complexity is minimized, and functionality optimized. For example, a single session of robot controlled proprioceptive training induces connectivity changes in the somatosensory networks associated with residual motor and sensory function [115], translating into improved motor function in stroke patients. This example perfectly illustrates the closed-loop epistemological link (Figure 6) between systems neuroscience and robotics/AI. Robotic sensory learning models inspired by self-organizing somatosensory network dynamics (in short: AI) are currently developed [115,116,117], within and well beyond the context of motor rehabilitation programs. Further studies on the effects of repeated training sessions on neural network learning and somatosensory functional plasticity will 1) enable the possible generalization of motor relearning and treatment effects in the clinical domain, and 2) the development of reliable robot motor learning and control on the basis of "strong" neuro-inspired AI.

**Funding:** This research received no external funding

**Acknowledgments:** The support of the CNRS is gratefully acknowledged

**Conflicts of Interest:** The authors declares no conflict of interest